\documentclass[a4paper,
               refpage       
               keeplastbox,   
               nospread,     
               ]{jacow}
\usepackage{pdfpages,multirow,ragged2e} %
\makeatletter%
	\ifboolexpr{bool{xetex}}
	 {\renewcommand{\Gin@extensions}{.pdf,%
	                    .png,.jpg,.bmp,.pict,.tif,.psd,.mac,.sga,.tga,.gif,%
	                    .eps,.ps,%
	                    }}{}
\makeatother

\ifboolexpr{bool{xetex} or bool{luatex}} 
 {}                                      
 {\usepackage[utf8]{inputenc}}           

\usepackage[USenglish]{babel}

\ifboolexpr{bool{jacowbiblatex}}%
 {%
  \addbibresource{jacow-test.bib}
  \addbibresource{biblatex-examples.bib}
 }{}
\begin{document}

\title{A $\mu$TCA-based Low Level RF System prototype for MYRRHA 100 MeV project}

\author{W. Sarlin, Y. Gargouri, C. Joly, J.-F. Yaniche, J. Lesrel, S. Berthelot, A. Escalda, M. Pereira, \\
		G. Ferry, N. Gandolfo, IPNO (UMR 8608 - CNRS/IN2P3), 91406 Orsay, France \\
		J. Bouvier, LPSC (UMR 5821 - CNRS/IN2P3), 38026 Grenoble, France \\
		P. Della Faille, SCK•CEN, 1348 Louvain-La-Neuve, Belgique}
	
\maketitle

\begin{abstract}
The first phase of the MYRRHA (Multi-purpose hYbrid Research Reactor for High-tech Applications) 100 MeV project, called MINERVA (MYRRHA Isotopes productioN coupling the linEar acceleRator to the Versatile proton target fAcility), was launched in September 2018. Through collaboration with SCK•CEN (Studiecentrum voor Kernenergie•Centre d'Étude de l'énergie Nucléaire), CNRS/IN2P3 (Centre National de la Recherche Scientifique/Institut National de Physique Nucléaire et de Physique des Particules) laboratories take in charge the developments of several parts of the future LINAC (LINear ACcelerator), including a fully equipped Spoke cryomodule prototype and a cold valves box. This cryomodule will integrate two superconducting single spoke cavities operating at 2K, the RF (Radio Frequency) power couplers and the cold tuning systems associated. In this context, a prototype of LLRF (Low-Level Radio Frequency), based on the $\mu$TCA (Micro Telecommunication Computing Architecture) standard, will be developed at IPNO (Institut de Physique Nucléaire d'Orsay). The main goal of this prototype is to design an innovative system that fulfils MYRRHA’s ambitious requirements (MTBF (Mean Time Between Failure) for the accelerator above 250 hours, restart in less than 3s in case of fault), as well as to provide a rugged design in prevision for the mass production phase. This paper presents the hardware and software architecture of this LLRF as well as the supervision through EPICS (Experimental Physics and Industrial Control System), and the status of these different tasks.
\end{abstract}

\section{INTRODUCTION}
The aim of the MYRRHA project [1] is to demonstrate radiotoxic waste transmutation through the design of a hybrid reactor, driven by a superconducting proton LINAC operating at 600 MeV. From the beginning of 2016, SCK•CEN is in charge of building the first section of the accelerator, up to 100 MeV (MINERVA project). As the R\&D phase started in July 2017, IN2P3 laboratories began the study of a SRF (Superconducting Radio Frequency) cryomodule (Beta=0.37, f=352.2 MHz) prototype. A complete test bench, aiming to conduct the full power RF tests planned for the second semester of 2021, will be accommodated by IPNO and housed in Orsay (France). IPNO is also in charge of the design and test of a $\mu$TCA-based [2] LLRF prototype. It benefits from the return on experience from the LLRF realized for MYRTE (MYRRHA Research and Transmutation Endeavour, design of an injector prototype) project [3].

\section{UPGRADE OF THE MYRTE LLRF}
The last tests campaign for the MYRTE LLRF occurred on September 2019. Regulation at the nominal power was performed, i.e. up to 110 kW, with good stability results [3]. Another approach was adopted to face MINERVA requirements, with a special view to reliability and mass production. Indeed, the latter implies a need for uniform, modular and rugged systems. Sustainability required for the facilities also lead to look for architecture with high availability and redundancy.

\begin{figure}[!htb]
   \centering
   \includegraphics*[width=\columnwidth]{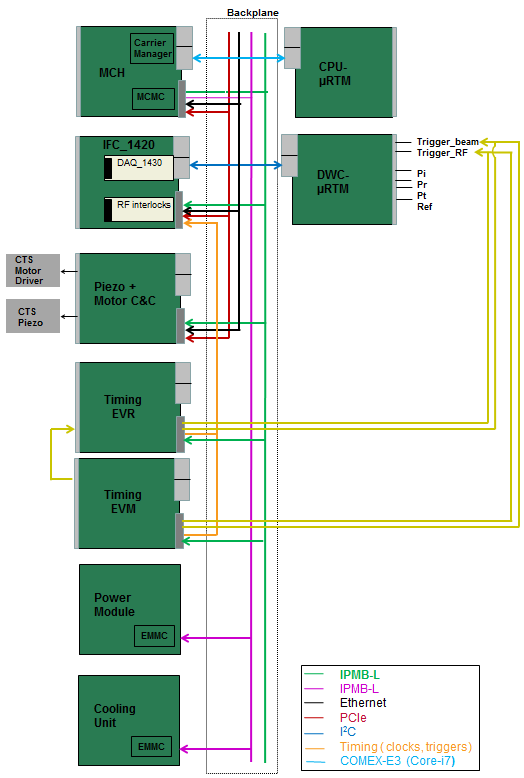}
   \caption{The IPNO $\mu$TCA-based LLRF for MINERVA.}
   \label{fig:fg1}
\end{figure}

Finally, the choice made at IPNO was to upgrade the existing system to a $\mu$TCA-based format, due to the high reliability provided by the standard (99.9\% without redundancy, 99.999\% else), as well as the number of on-the-shell industrial solutions for each component of a LLRF. The complete system is presented in the next section, with the different electronical boards and their dedicated functions.

\section{A $\mu$TCA VERSION OF THE LLRF }

The $\mu$TCA-based LLRF under development at IPNO follows the architecture presented in Fig. 1.

Currently, all the hardware mentioned, except the down-converter $\mu$RTM (Micro Rear Transition Module) module and the multipacting board (embedded in the IFC\_1420), has been acquired and assembled in a Native-R5 chassis from N.A.T. (Gesellschaft für Netzwerk- und Automatisierungs-Technologie mbH).

\subsection{Hardware of the MINERVA LLRF}
\vspace{0.2cm}

The different electronical boards that compose the LLRF as presented in Fig. 1 are:
\begin{Itemize}
    \item IFC\_1420: For the digital part of LLRF acquisition, we selected the IFC-1420 Fast Digitizer AMC (Advanced Mezzanine Card) that has been developed by IOxOS Technologies. This AMC is an analog variant of the IFC\_1410 with an onboard ADC/DAC function (DAQ-1430) that includes two FPGA (Field-Programmable Gate Array) Mezzanine Card (FMC) slots:
\begin{Itemize}
    \item DAQ\_1430: Mezzanine board for data acquisition. It implements ten 16-bit ADC channels at 250 Msps and four 16-bit DAC channels at 2.5 Gsps dedicated to the $\mu$RTM interface.
    \item RF Interlocks: a multipacting and arc detection board dedicated to the RF couplers with several inputs and outputs for externals interlocks from the cryogenic and auxiliaries PLC (Programmable Logic Controller) in particular and LLRF status.                 
\end{Itemize}

These FMCs are both controlled by a Xilinx Kintex UltraScale FPGA KU040. The IFC-1420 board embeds a PowerPC (NXP QorIQ T2081 processor @ 1.8 GHz) CPU (Central Processing Unit) and supports a $\mu$RTM (Micro Rear Transition Module) card compliant to DESY A1.1CO to supply the analog signals to the ADC function.
\break
    \item Down-converter: Off-the-shelf RF board for both LLRF and diagnostics systems, as beam position monitor, for instance. It will operate for all accelerating sections frequencies, in order to use the same VHDL (VHSIC (Very High Speed Integrated Circuits) Hardware Description Language) code for each (no specific code at low frequency due to direct sampling operation). It is based on five RF channels (50 ohms) in the frequency range 80MHz to 1000MHz with a down conversion to an intermediate frequency range from 7.5 to 50MHz. The Local Oscillator (LO) signal will be provided by a low phase noise Phase-Locked Loop (PLL) tuned to the accelerator frequency. It is developed as a $\mu$RTM module in MTCA.4 and will be connected directly to the IFC\_1420. The main challenge is to implement all functionalities with very low crosstalk between channels or temperature stability, amongst other requirements, with a usable area close to 236 cm² (one side of a full-size double module). The design of this board is currently under development through a collaboration between the Electronics for Accelerators Group at IPNO and IOxOS Technologies, within the framework of a Non-Disclosure Agreement. The first prototype is planned for the next year.
\break
    \item Piezo + Motor C\&C: This AMC board associated to an external linear power rack allows the frequency tuning of a cavity using piezoelectric and motor controllers. A first prototype including only the piezoelectric controller has been developed and validated through collaboration between IPNO and LPSC (Laboratoire de Physique Subatomique et de Cosmologie).
\break
    \item Timing EVM (EVent Master): The mTCA-EVM-300 board from MRF (Micro Research Finland) is used for generating and distributing the RF and beam triggers with event codes transmitted by optical fiber to the EVRs.
\break
    \item Timing EVR (EVent Receiver): The mTCA-EVR-300U board, also from MRF, decodes events and signals from the EVM, and distributes the RF and beam triggers to the system.
\break
\vspace{0cm}
    \item MCH ( $\mu$TCA Carrier Hub): This crucial functionality, that provides a centralized management of the whole µTCA system, is performed by a NAT-MCH-PHYS board, from N.A.T. It is also equipped with a $\mu$RTM module providing a CPU for general purposes, consisting of a NAT-MCH-RTM-COMex-i7 board.
\break
    \item Power module: realized by a NAT-PM-AC600D board from N.A.T (110-240VAC, 600W output).
\break
    \item Cooling unit: 4 fans for AMCs side and 2 for RTMs side, integrated directly into the crate.
\end{Itemize}

\vspace{0.2cm}
\subsection{FPGA and CPU description system using VHDL and EPICS language}
\vspace{0.2cm}

\begin{figure}[!htb]
   \centering
   \includegraphics*[width=\columnwidth]{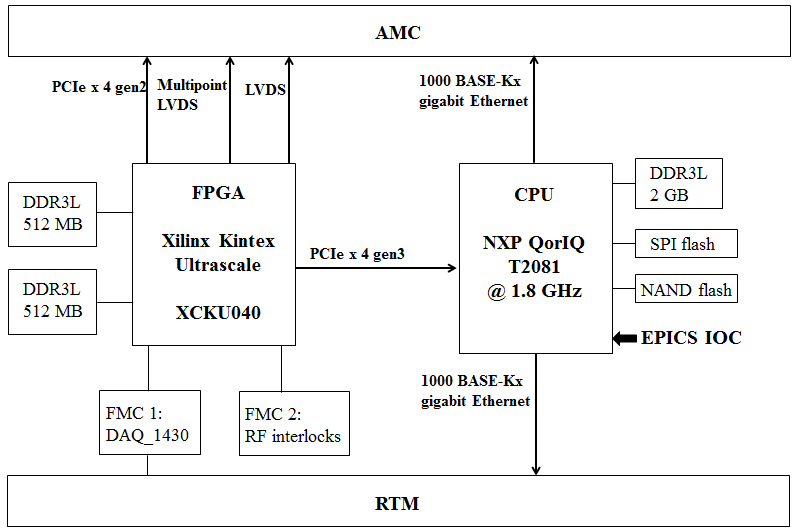}
   \caption{Memories and connections of FPGA and CPU.}
   \label{fig:fg2}
\end{figure}

Figure 2 illustrates the different memories directly associated to the FPGA and the CPU as well as connections that allow communication between the FPGA and the CPU, the AMC (IFC\_1420) and the $\mu$RTM module (Down-converter). The FPGA is connected to two 512 MByte DDR3L-1066 SDRAM memories, and to the AMC interface with two PCIe x4 Gen2 links, point-to-point links and multi-point links. It is also connected to the processor with a PCIe x4 Gen3 link.

The processor itself is connected to 2 GByte DDR3L-1866 SDRAM, 512 MByte NAND flash (non-volatile storage memory) and 64 MByte SPI memory (non-volatile boot memory). It is connected by one 1000BASE-KX gigabit Ethernet link to the AMC interface and one 1000BASE-KX link to the RTM interface.

\section{VHDL PROGRAMMING USING TOSCA IIb}
We will present in this paragraph the necessary FPGA functions, which will communicate with the LLRF control system, their location in the FPGA and how to interface them by exploiting TOSCA IIb architecture.
\vspace{0.2cm}
\subsection{TOSCA IIb }
\vspace{0.2cm}
In the IFC-1420 digitizer, the FPGA processing unit is powered by IOxOS Technologies proprietary TOSCA FPGA Design Kit, a system design environment for custom application design and integration.

TOSCA is a NoC (Network on Chip) architecture based on a PCI Express switch centric structure that implements a memory mapped model with segregated I/O Space (CONTROL Plane) and Memory Space (DATA Plane) [4]. 

The version of TOSCA used for current developments in Vivado is TOSCA IIb.
\vspace{0.2cm}
\hyphenation{system} 
\subsection{FPGA functions interacting with the control system}
\vspace{0.2cm}
FPGA functionalities are divided into three categories:
\break
\begin{Itemize}
    \item Set-points and Boolean indicators (FPGA register) used for alarms, command and configuration  parameters, Boolean states, etc...
\break
    \item FIFO (First In, First Out) memory: Medium-sized Frame (e.g. for data acquisition) and pattern signals.
\break
    \item DMA (Direct Access Memory): monitoring frame and circular buffer.             
\end{Itemize}
\vspace{0.2cm}
\subsection{TOSCA interfaces}
\vspace{0.2cm}
Several interfaces have been implemented in TOSCA IIB environment. Among them, we selected:
\break
\begin{Itemize}
    \item TCSR interface to implement control and status registers. TCSR Bus interfaces support single beat Read/Write transactions, which allows the 32-bit data from the FPGA to be send to the supervision and vice versa.
\break
    \item TMEM (Tosca MEMory) interface, which is 64-bit wide. It supports single beat and burst in Read/Write operations. TMEM will allow communicating FIFOs with the control system.
\break
    \item SMEM (Shared MEMory) interface, a high speed 64-bit interface that supports two independent Read/Write transactions. It is used in particular to manage large data buffer integrated in SMEM DDR3L. 
\break
    \item iDMA interface: it is used to send 64-bit data via DMA. This interface uses two independent engines (Read\_Engine and Write\_Engine).
\break
\end{Itemize}

Table 1 summarizes the FPGA’s functions that communicate with the control system, the corresponding TOSCA interfaces and their data size. It also presents a primary study of the number of elements that will be used for each functionality based on MYRTE project data.

\begin{table}[!hbt]
   \centering
   \begin{center}
   \caption{The different FPGA functionalities with the corresponding TOSCA IIb memories (*: MYRTE feedback).}
   \resizebox{\columnwidth}{!}{%
   \begin{tabular}{>{\centering\arraybackslash}m{2cm} >{\centering\arraybackslash}m{2cm} >{\centering\arraybackslash}m{2cm} >{\centering\arraybackslash}m{1cm} >{\centering\arraybackslash}m{2cm}}
       \toprule
       \textbf{Function} & \textbf{FPGA Location} & \textbf{TOSCA Interface} & \textbf{Data Size} & \textbf{Number of  Elements}\\
       \midrule
          Set-points, Boolean indicators & Registers & TCSR & 32 bits & ~40*        \\ 
         \hline
          Medium Frame & FIFOs & TMEM & 64 bits & 10*        \\ 
         \hline         
          Patterns & FIFOs & TMEM  /SMEM & 64 bits & 2*        \\ 
          \hline 
          Monitoring Frame & DMA & iDMA & 64 bits & 1*        \\
          \hline
          Circular Buffer (Simple/Double) & DMA & iDMA & 64 bits & 1-2        \\
       \bottomrule
   \end{tabular}
   }
   \end{center}
   \label{tab:tb1}
\end{table}

So far, the different memories have been investigated and are already used, except for the iDMA interface. Once finished, the next step will be to integrate the signal processing developments realized for MYRTE, which will constitute a starting basis for VHDL developments [3]. Nevertheless, feedback from MYRTE raises the need for upgrades to improve the latency of the system (to increase gain and phase margin of the LLRF), the signal-to-noise ratio (use of near I/Q demodulation instead of I/Q), amongst others.
\section{EPICS}
\subsection{ESS EPICS Environment }
\vspace{0.2cm}
For MINERVA, EPICS was chosen to perform control and command. Developments made so far at IPNO followed the model of E3 (ESS EPICS Environment), the control architecture used by ESS (European Spallation Source). Using the open sources from the project [5], a complete architecture was deployed on the $\mu$TCA crate, thus the developing work only has to focus on the design of two E3-compliant modules: one for the general supervision of the crate, and another one for the LLRF.
\vspace{0.2cm}
\subsection{MINERVA $\mu$TCA Health Monitoring IOC}
\vspace{0.2cm}
After installing E3 from the sources [5], one module included in the architecture, called ipmi-epics, served as a basis for the design of a general supervision IOC (Input-Output Controller). ipmi-epics is mainly a wrapper library to integrate IPMItools [6] into EPICS. Basically, ipmi-epics can be used to scan the system through IPMI (Intelligent Platform Management Interface) bus communications, and generate PVs (Process Variables) that correspond to each detected sensors in the system, for instance temperature, voltage or current sensors. This was used at IPNO to design an IOC (Input-Output Controller) dedicated for the general supervision of the crate (Fig. 3).

\begin{figure}[!htb]
   \centering
   \includegraphics*[width=\columnwidth]{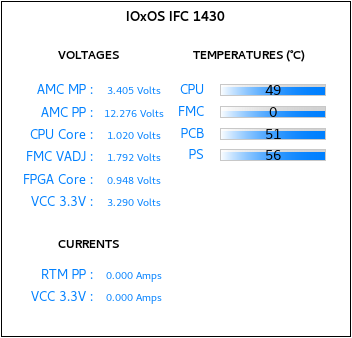}
   \caption{Example of collected information with the modified ipmi-epics module for the IOxOS 1430 board.}
   \label{fig:fg3}
\end{figure}

However, one limitation to this process was found for a complete setup: with different boards present, it was not possible to relate one PV to its physical hardware, and moreover it was not possible to ensure PV’s name unicity. Therefore, a new version of the E3 module was implemented at IPNO, which relies on FRU (Field Replaceable Unit) data to characterize each pieces of hardware through the use of two numbers: a generic identifier and an instance number. Then, during the IPMI scan for sensor detections, these two numbers were again found in the sensor’s field, forming a couple \{ID,Instance\}: it was thus possible to establish a link between each sensor and its dedicated hardware (FRU device). This easily led to the unicity for each PV’s name generated. Thus, according to the MYRRHA naming convention [7], variables were defined with the following naming scheme: 

\begin{center}
PREFIX:BOARD\_NAME\_BOARD\_ID\_PV\_NAME \\
PREFIX=SYSTEM-SUBSYSTEM:DEVICE\_TYPE-DEVICE\_INSTANCE
\end{center}

In addition, the collection of the triplet \{Manufacturer,Product,Serial N°\} appears to be a great help to maintenance operations, as each hardware component is now fully characterized in EPICS (Fig. 4) and can quickly be isolated thanks to its serial number in case of failure.

\begin{figure}[!htb]
   \centering
   \includegraphics*[width=\columnwidth]{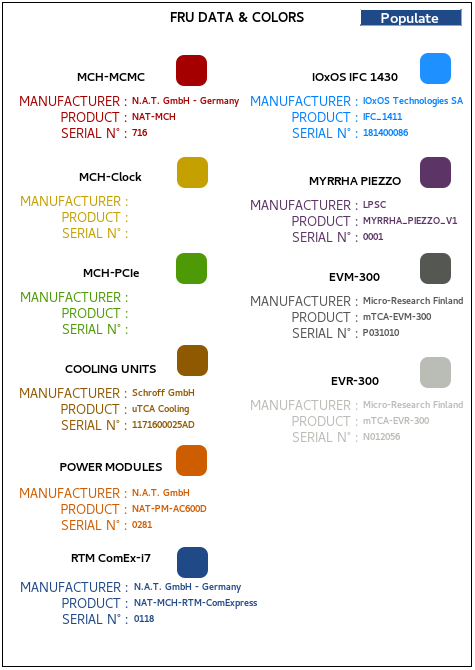}
   \caption{FRU data for each board on the systems, showing the triplet \{Manufacturer, Product, Serial N°\}.}
   \label{fig:fg4}
\end{figure}

\subsection{Tscmon Lib Integration}
\vspace{0.2cm}
Last but not least are the developments of a E3-compliant module for the control and command of the LLRF. To this end, work is under progress to integrate the TOSCA IIb C library, toscamon, into EPICS. This software utility, provided by IOxOS Technologies, is indeed fully interfaced with the FPGA and was successfully tested in read/write modes with the TCSR registers and TMEM memories. This integration constitutes the main remaining task for the EPICS developments, as feedback from MYRTE already provides the majority of the signal processing architecture [3].  

\section{CONCLUSION}

Within the framework of MINERVA, a $\mu$TCA-based LLRF prototype is being developed at IPNO, taking advantage of MYRTE feedback as well as the possibilities given by the $\mu$TCA standard. The first step of the design, the selection of the different hardware components and the assembly of a complete $\mu$TCA crate, was finished by the end of 2018. Special attention was given to mass production and reliability considerations while choosing the hardware, the only in-house developments being the down-converter in collaboration with IOxOS Technologies and the RF interlocks FMC. The following phase consisted on the deployment of E3 and the design of the $\mu$TCA health monitoring IOC, and the mastering of the TOSCA IIb environment provided by IOxOS Technologies. 

Remaining work consists of both the integration of MYRTE VHDL developments for signal processing into the TOSCA environment, and the design of a wrapper module to integrate TOSCA C library into E3. The design of this new version for the LLRF is under progress, and a first prototype is planned for 2020, so that it could be validated through on-site tests, before the RF tests taking place during the second semester of 2021.

\section{ACKNOWLEDGEMENT}
The presented work has been funded and was made possible by the collaboration between SCK•CEN and the IN2P3 (CNRS).

The authors would like to thank IOxOS Technologies team for their technical support on TOSCA IIb as well as for the ongoing collaboration for the $\mu$RTM down-converter. 

The authors also thank Olivier Bourion (LPSC) for the support provided for IPMItool understanding, and the work on the piezo board. 


%
%
\ifboolexpr{bool{jacowbiblatex}}%
	{\printbibliography}%
	{%
	
	
} 
%
%

\null
\end{document}